\documentstyle[nato,epsfig,pstricks,amstext]{crckapb} 

\begin{opening}

\title{TRUNCATED OVERLAP FERMIONS: THE LINK BETWEEN OVERLAP
AND DOMAIN WALL FERMIONS}

\author{Artan Borici}

\institute{Paul Scherrer Institute \\
           CH-5232 Villigen PSI}

\end{opening}

\runningtitle{TRUNCATED OVERLAP FERMIONS}

\begin{document}

\begin{abstract}
In this talk I will emphasize the role of the Truncated Overlap Fermions
in showing the equivalence between the Domain Wall and Overlap Fermions
up to an irrelevant factor in the fermionic integration measure.
I will also show how Domain Wall type fermions with a finite number of flavors
can be used to accelerate propagator calculations of their light partner
in the infinite flavor limit.
\end{abstract}

\section{Introduction}

It required some time until Domain Wall \cite{Kaplan,Shamir&Furman_Shamir}
and Overlap \cite{Narayanan_Neuberger} formulations of chiral lattice fermions
gained the adequate momentum \cite{Luscher_Lat99&Neuberger_Lat99}.
A remnant chiral symmetry on the lattice,
called the Ginsparg-Wilson relation \cite{Ginsparg_Wilson},
that was more recently noticed \cite{Hasenfratz_Laliena_Niedermayer,Luscher1},
was shown to be the building block of a chiral gauge theory which exists
on the lattice \cite{Luscher2}.

The basic idea of Domain Wall Fermions is an expanded flavor space which may be seen
as an extra dimension with 
left and right handed fermions defined in the two opposite
boundaries or walls, as it is sketched schematically below.

Let $N$ be the size of the extra dimension, $D_W$ the Wilson-Dirac
operator, and $m$ the bare fermion mass. Then, the theory with
{\it Domain Wall Fermions} is defined by the action
\cite{Kaplan,Shamir&Furman_Shamir}: 
\begin{equation}
S_{DW} := \bar{\Psi} a_5{\cal M}_{DW} \Psi = \sum_{i=1}^{N}
  \bar{\psi}_i [(a_5D^{||}-1)\psi_i + P_{+}\psi_{i+1} + P_{-}\psi_{i-1}]
\end{equation}
with boundary conditions given by
\begin{equation}
\begin{array}{l}
P_{+}(\psi_{N+1} + m \psi_1) = 0 \\
P_{-}(\psi_0 + m \psi_N) = 0
\end{array}
\end{equation}
where $\cal{M}$ is the five-dimensional fermion matrix of the
regularized theory and
$D^{||} = M - D_W$ with $M \in (0,2)$ being a mass parameter
and $a_5$ the lattice spacing in the $5^{th}$ direction.

\vspace{3cm}
\pspolygon[linewidth=1pt](0,0)(0,2)(1,2.5)(1,.5)
\pspolygon[fillstyle=hlines](0,0)(0,2)(1,2.5)(1,.5)
\pspolygon[linewidth=1pt](2,0)(2,2)(3,2.5)(3,.5)
\rput(4,1){BULK}
\psdots(5,1)(6,1)(7,1)
\pspolygon[linewidth=1pt](10,0)(10,2)(11,2.5)(11,.5)
\pspolygon[fillstyle=hlines](10,0)(10,2)(11,2.5)(11,.5)
\psline[linewidth=2pt]{->}(3,-1)(7,-1)
\rput(5,-0.5){$5^{th}$ dimension}
\vspace{2cm}

The theory with {\it Truncated Overlap Fermions}
is defined by \cite{Borici_TOV,Borici_MG}:
\begin{equation}
S_{DW} := \bar{\Psi} a_5{\cal M}_{TOV} \Psi = \sum_{i=1}^{N}
  \bar{\psi}_i [(a_5D^{||}-1)\psi_i
  + (a_5D^{||}+1)P_{+}\psi_{i+1} + (a_5D^{||}+1)P_{-}\psi_{i-1}]
\end{equation}
In both cases the lattice spacing $a$ of the four dimensional
theory is set to one.

Truncated Overlap Fermions can be formally constructed from
Domain Wall Fermions by substituting
\begin{equation}
\begin{array}{l}
P_{+}\psi_{i+1} \rightarrow (a_5D^{||} + 1)P_{+}\psi_{i+1} \\
P_{-}\psi_{i-1} \rightarrow (a_5D^{||} + 1)P_{-}\psi_{i-1}
\end{array}
\end{equation}
while the boundary conditions remain the same as before.

\section{Continuum limit in the $5^{th}$ dimension}

Let me first write down the operator kernels of both theories:
\begin{equation}
\begin{array}{l}
{\cal M}_{DW} = D^{||} + \frac{1}{a_5} (e^{a_5 \gamma_5 \partial_5} - 1) \\
{\cal M}_{TOV} = D^{||} (e^{a_5 \gamma_5 \partial_5} + 1)
 + \frac{1}{a_5} (e^{a_5 \gamma_5 \partial_5} - 1)
\end{array}
\end{equation}
This form can be easily checked by using the identity:
\begin{equation}
\begin{array}{l}
\psi(t_5+a_5) = e^{a_5 \partial_5} \psi(t_5) \\
\text{with} ~~\psi_i \equiv \psi(t_5 = i a_5)
\end{array}
\end{equation}
Taking the limit $a_5 \rightarrow 0$, I get:
\begin{equation}
\begin{array}{l}
{\cal M}_{DW} = D^{||} + \gamma_5 \partial_5 + \frac{a_5}{2} \partial_5^2 \\
{\cal M}_{TOV} = D^{||} (2 + a_5 \gamma_5 \partial_5)
 + \gamma_5 \partial_5 + \frac{a_5}{2} \partial_5^2
\end{array}
\end{equation}

Hence, in a continuous flavor space, both theories are unique up to
an asymmetric factor remaining from the Truncated Overlap. Therefore,
one may conclude that Domain Wall and Truncated Overlap Fermions are
discretizations of the same Domain Wall fermion theory in the continuous
flavor space $[0,T_5]$:

\vspace{3cm}
\pspolygon[linewidth=1pt](4.5,0)(4.5,2)(5.5,2.5)(5.5,.5)
\pspolygon[linewidth=1pt](8,0)(8,2)(9,2.5)(9,.5)
\psline[linewidth=2pt]{|-|}(3,1.2)(11,1.2)
\rput(2.8,0.8){0}
\rput(11.4,0.8){$T_5$}

\noindent
defined by the following action:
\begin{equation}
S = \Psi (\sl{D} + \gamma_5 \partial_5 - M) \Psi \\
\end{equation}
with the following boundary conditions:
\begin{equation}
\begin{array}{l}
P_{+} [\Psi(\cdot,T_5) + m \Psi(\cdot,0)] = 0 \\
P_{-} [\Psi(\cdot,0) + m \Psi(\cdot,T_5)] = 0 \\
\end{array}
\end{equation}

\section{Ginsparg-Wilson relation}

A remnant chiral symmetry on the lattice may be possible
if one allows a local symmetry breaking for propagating states.
This statement is encoded in the
Ginsparg-Wilson relation \cite{Ginsparg_Wilson}:
\begin{equation}\label{GWR}
\gamma_5 D^{-1} + D^{-1} \gamma_5 = 2a \gamma_5 R,
\end{equation}
where $D$ is a local Dirac operator and
$R$ is also a local operator trivial in Dirac space.

An explicit solution of this relation is given by the Overlap
Dirac operator \cite{Neuberger1}.
In fact one can show that a Dirac operator obeying the
Ginsparg-Wilson symmetry can be derived
from the Domain Wall \cite{Neuberger_DWF&Kikukawa_Noguchi,Borici_TOV},
and Truncated Overlap Fermions \cite{Borici_TOV} in the infinite
flavor limit.

The situation is unclear when the number of flavors is finite \cite{Borici_TOV}.
I present here some preliminary tests of the Ginsparg-Wilson relation
on a small number of configurations on a $4^4$ lattice at $\beta = 6$.

In Figs. 1-2 the locality of the Dirac operator is observed for $N=4$ and
$N=32$ number of flavors. The behavior of $R$ is tested in Figs. 3-4.
These suggest that $R$ tends towards a Kronecker-Delta function as the
number of flavors grows and the convergence is faster
for Domain Wall Fermions. More data are needed to verify this evidence.

\section{Infinitely separated walls}

The results of the previous section, although preliminary, are enough to
conclude that the infinite limit in
the fifth dimension is needed. This may be unrealistic for practical
computations, if one would keep working with the whole $5-$dimensional
theory.

A simple solution is to work in the four dimensional framework
of the Overlap Dirac operator \cite{Neuberger1}:
\begin{equation}
D = \frac{1+m}{2} - \frac{1-m}{2} \gamma_5 \text{sgn}(H)
\end{equation}
where the Hamiltonian $H = \gamma_5 D^{||}$ corresponds to the
``evolution'' in the fifth dimension of Truncated Overlap Fermions with
a transfer matrix given by \cite{Borici_TOV}:
\begin{equation}
T_{TOV} = \frac{1 + H}{1 - H}
\end{equation}

For Domain Wall Fermions it is not straightforward to construct ``easy to use''
Hamiltonians, since the transfer matrix is given by \cite{Borici_TOV}:
\begin{equation}
T_{DW} = \frac{1}{1 + H P_{-}}(1 - H P_{+})
\end{equation}
where numerator and denominator do not commute.

In analogy to Truncated Overlap Fermions, I define a Hamiltonian $\cal H$ for
Domain Wall Fermions, such that the transfer matrices of both theories coincide:
\begin{equation}
\frac{1 + {\cal H}}{1 - {\cal H}} = \frac{1}{1 + H P_{-}}(1 - H P_{+})
\end{equation}
from which I can write down the solution:
\begin{equation}
{\cal H} = \gamma_5 \frac{D^{||}}{2 - D^{||}} = H \frac{1}{2 - D^{||}}
\end{equation}
where $a_5 = 1$ is assumed.

This looks merely a trick, but in fact  it is obvious by the definition that
$\cal H$ derives from the transfer matrix of the
Domain Wall Fermions. Therefore, I arrive to the conclusion that

\noindent
{\em The light fermion operator in the infinite flavor limit
of Domain Wall Fermions is given by the Overlap Dirac operator
with Hamiltonian $\cal H$.}

Some remarks are in order here:

a) The form of $\cal H$ suggests that both theories
are identical in the limit $a \rightarrow 0$. In this case ${\cal H} \approx H$.

b) For finite $a$ any theory with Wilson fermions can be equivalently defined
to a theory with a Dirac operator:
\begin{equation}
\begin{array}{l}
\frac{D^{||}}{2 - D^{||}} = \frac{D_W}{1 + D_W} \\
 \text{for} ~~M = 1 ~~\text{and} ~~m = 0
\end{array}
\end{equation}
up to the determinant factor det$(1 + D_W)$. This is easily seen by the identity:
\begin{equation}
\int_{\psi \bar{\psi}} e^{- (\bar{\chi} - \bar{\psi})(\chi - \psi)
 - \bar{\psi} D_W \psi} = \text{det}(1 + D_W) e^{- \bar{\chi} \frac{D_W}{1 + D_W} \chi}
\end{equation}
i.e. the new operator is the Schur complement of the
new ``effective'' theory with free
fermions $\chi, \bar{\chi}$. Therefore up to an irrelevant determinant factor,
both theories are equivalent for finite $a$.

{\em Computational remarks on  $\cal H$.}

It is important to know the computation overhead of $\cal H$ if one would like
to work with Domain Wall Fermions in the infinite flavor limit, i.e. in the
Overlap framework.

Practical methods to compute the Overlap operator use the application of
$\cal H$ or ${\cal H}^2$ to a vector
\cite{Borici_OV,Neuberger2,Edwards_Heller_Narayanan}.

It is obvious that the
computation of $\cal H$ is more complex than that of $H$,
although the inversion of
$2 - D^{||}$ is well conditioned and can be done fast.

On the other hand,
$\cal H$ is conditioned better than $H$.
To illustrate this, I have computed the spectrum of $D^{||}/(2 - D^{||})$ for
free fermions on a $16^4$ lattice and also for
a fixed background at $\beta = 6$ on a $4^4$ lattice. The spectra are shown
in Figs. 5-7.

\section{Inversion of the Overlap Dirac operator}

It has been pointed out that Truncated Overlap Fermions
can be used to compute efficiently the inverse of the Overlap Dirac operator
\cite{Borici_TOV,Borici_MG}.
From the discussion above, it can be concluded that Domain Wall Fermions
can also be used effectively to compute the propagation of the light
fermion in the infinite flavor limit.

The basic idea is a multigrid algorithm, which is illustrated below.

\vspace{2cm}
\psline[linewidth=2pt]{|-|}(3,1.2)(11,1.2)
\rput(7,1.6){$N = \infty$}
\rput(2.8,0.8){0}
\rput(11.4,0.8){$T_5$}
\psline[linewidth=2pt]{<->}(7,0.8)(7,-0.2)
\psline[linewidth=2pt]{|-|}(3,-0.8)(11,-0.8)
\psdots[dotstyle=|](4,-0.8)(5,-0.8)(6,-0.8)(7,-0.8)(8,-0.8)(9,-0.8)(10,-0.8)
\rput(2.8,-1.2){1}
\rput(11.4,-1.2){$N$}
\vspace{2cm}

I would like to solve the linear system:
\begin{equation}
D z = b
\end{equation}
where $D$ is the chiral Dirac operator and $z_0$ is a first guess.
The algorithm
may be described as a three step iteration scheme \cite{Borici_TOV,Borici_MG}:

I. Compute the quark propagator in the Domain Wall framework, i.e. finite $N$,
which can be interpreted as a coarse lattice propagator:
\begin{equation}
\begin{array}{l}
{\cal M} P x_5 = {\cal M}_1 P b_5 \\
b_5 = (b,0,\ldots,0)^T \\
x_5 = (x,y,\ldots,z)^T 
\end{array}
\end{equation}
where $x_5$ and $b_5$ are block-vectors with $N$ blocks,
${\cal M}_1$ is the same matrix $\cal M$ but with bare quark mass
$m = 1$, and $P$ is the
following permutation operator:
\begin{equation}
\begin{array}{l}
(P x_5)_i = P_{+} (x_5)_i + P_{-} (x_5)_{i+1}, ~~i = 1, \ldots, N-1 \\
(P x_5)_N = P_{+} (x_5)_N + P_{-} (x_5)_1
\end{array}
\end{equation}

II. Compute the residual error in the Overlap framework:
\begin{equation}
\begin{array}{l}
z = z_0 + x \\
r = b - D z
\end{array}
\end{equation}

III. Construct the new residual error of the five dimensional theory and define the
new approximate solution:
\begin{equation}
\begin{array}{l}
b_5 \leftarrow (r,0,\ldots,0)^T \\
z_0 \leftarrow z
\end{array}
\end{equation}
and go to step I. or otherwise stop.

The scheme is tested on $30$ small $4^4$ lattices at $\beta = 6$ for
the Overlap Fermions. The results
are shown in Fig. 8, where the multigrid pattern of the residual norm is clear.
For comparison, in Fig. 8 are shown
the results of directly applying the Conjugate Residuals
(CR) algorithm. The gain is about a factor $10$ in this case. More results are
needed on larger lattices.

Note that CR is the best, i.e. the optimal algorithm for the Overlap operator,
which is a normal operator \cite{Borici_thesis}. To invert the ``big'' matrix
in step I., I have used the BiCGstab2 algorithm \cite{MGutknecht} which is almost
optimal in most of the cases for the non-normal matrices as its is the matrix
$\cal M$  \cite{Borici_thesis}.

\section{Conclusions}

I have shown the equivalence between Domain Wall and Overlap Fermions up
to an irrelevant factor in the fermionic integration measure.

Domain Wall and Truncated Overlap Fermions can be used to accelerate the
computation of wall propagators in the infinite flavor limit.

\section{Acknowledgements}

The author thanks PSI where part of this work was done and the CSCS for
the allocation of the computing time on the NEC/SX4.

The author would like to thank Valentin Mitrjushkin for the invitation and
kind hospitality at this Workshop.

\pagebreak

\begin{figure}
\epsfxsize=8cm
\vspace{3cm}
\centerline{\epsffile[100 200 500 450]{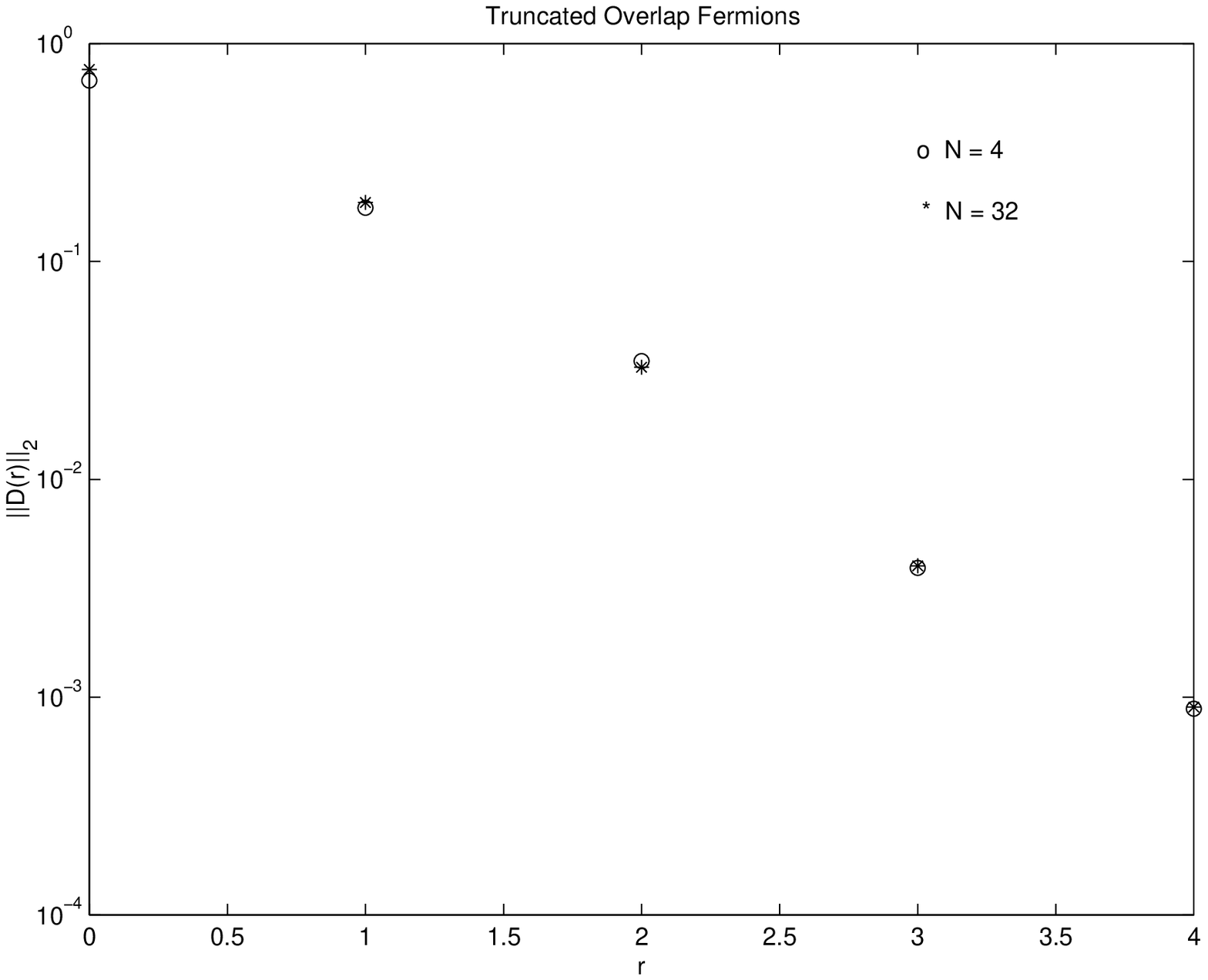}}
\caption{
Norm of the $D$ kernel in spin and color space with the distance $r$ from
the origin for $N = 4$ (circles) and $N = 32$ (stars).
}
\end{figure}
\begin{figure}
\epsfxsize=8cm
\vspace{3cm}
\centerline{\epsffile[100 200 500 450]{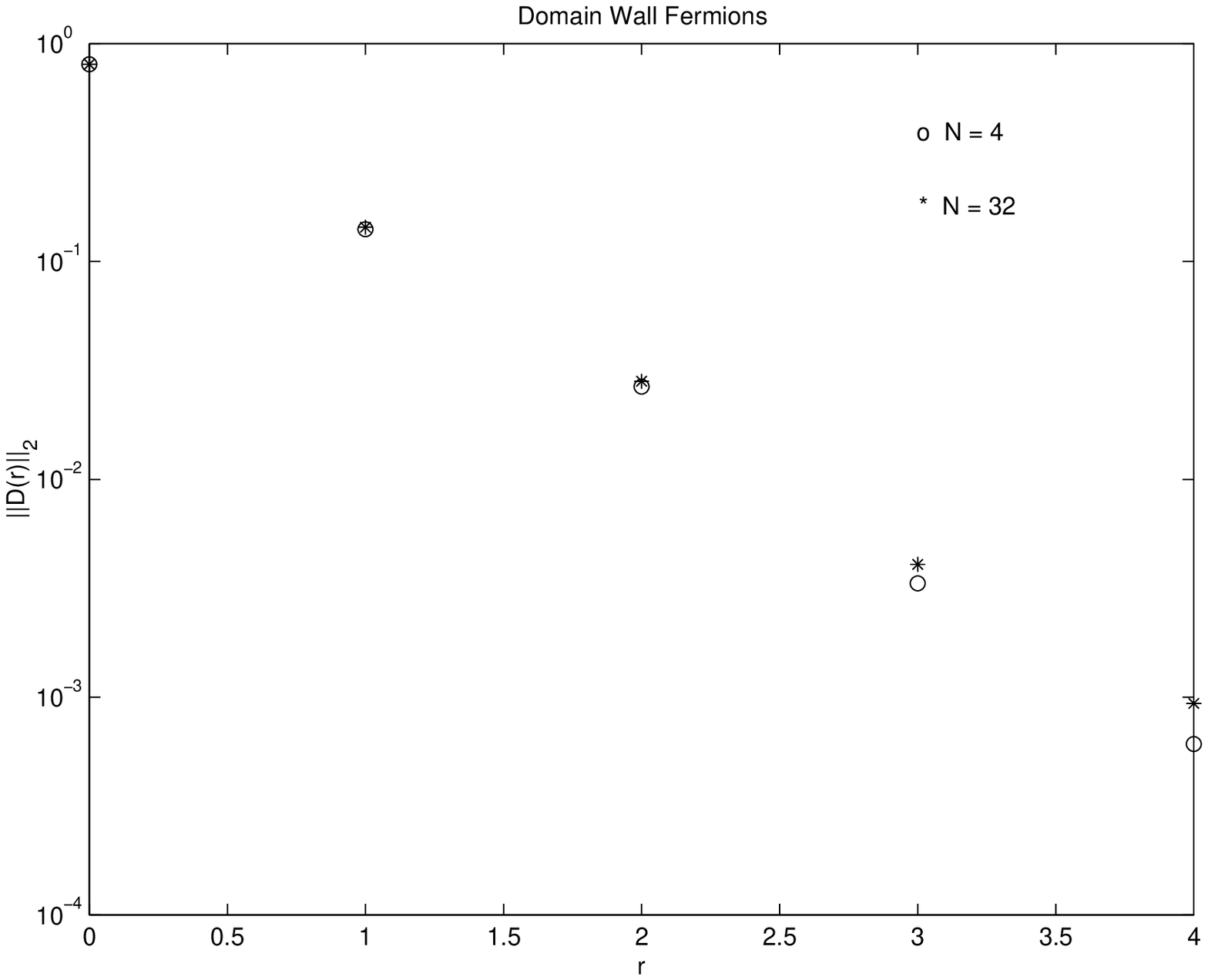}}
\caption{
Norm of the $D$ kernel in spin and color space with the distance $r$ from
the origin for $N = 4$ (circles) and $N = 32$ (stars).
}
\end{figure}
\begin{figure}
\epsfxsize=8cm
\vspace{3cm}
\centerline{\epsffile[100 200 500 450]{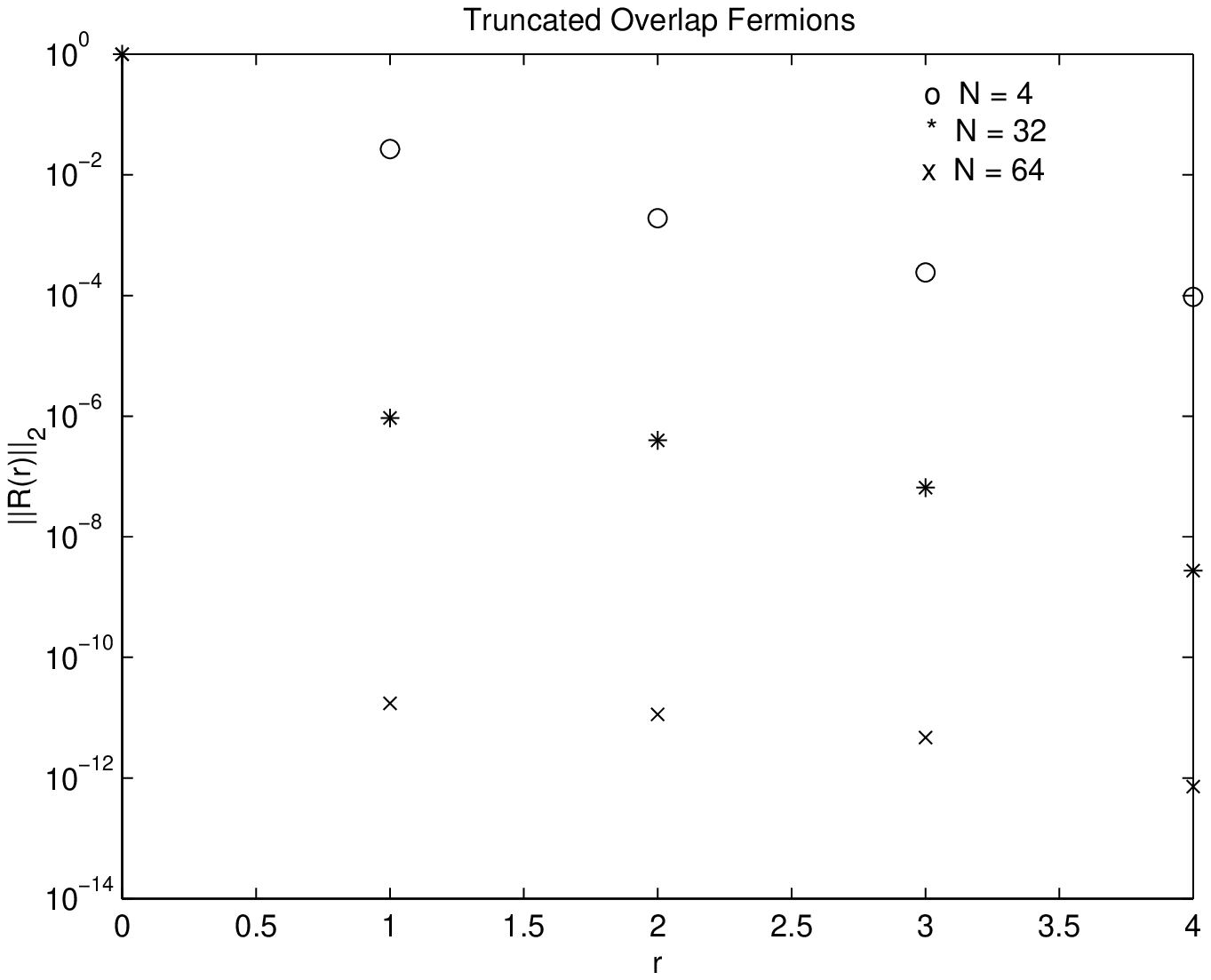}}
\caption{
Norm of the $R$ kernel in spin and color space with the distance $r$ from
the origin for $N = 4$ (circles), $N = 32$ (stars) and $N = 64$ (crosses).
}
\end{figure}
\begin{figure}
\epsfxsize=8cm
\vspace{3cm}
\centerline{\epsffile[100 200 500 450]{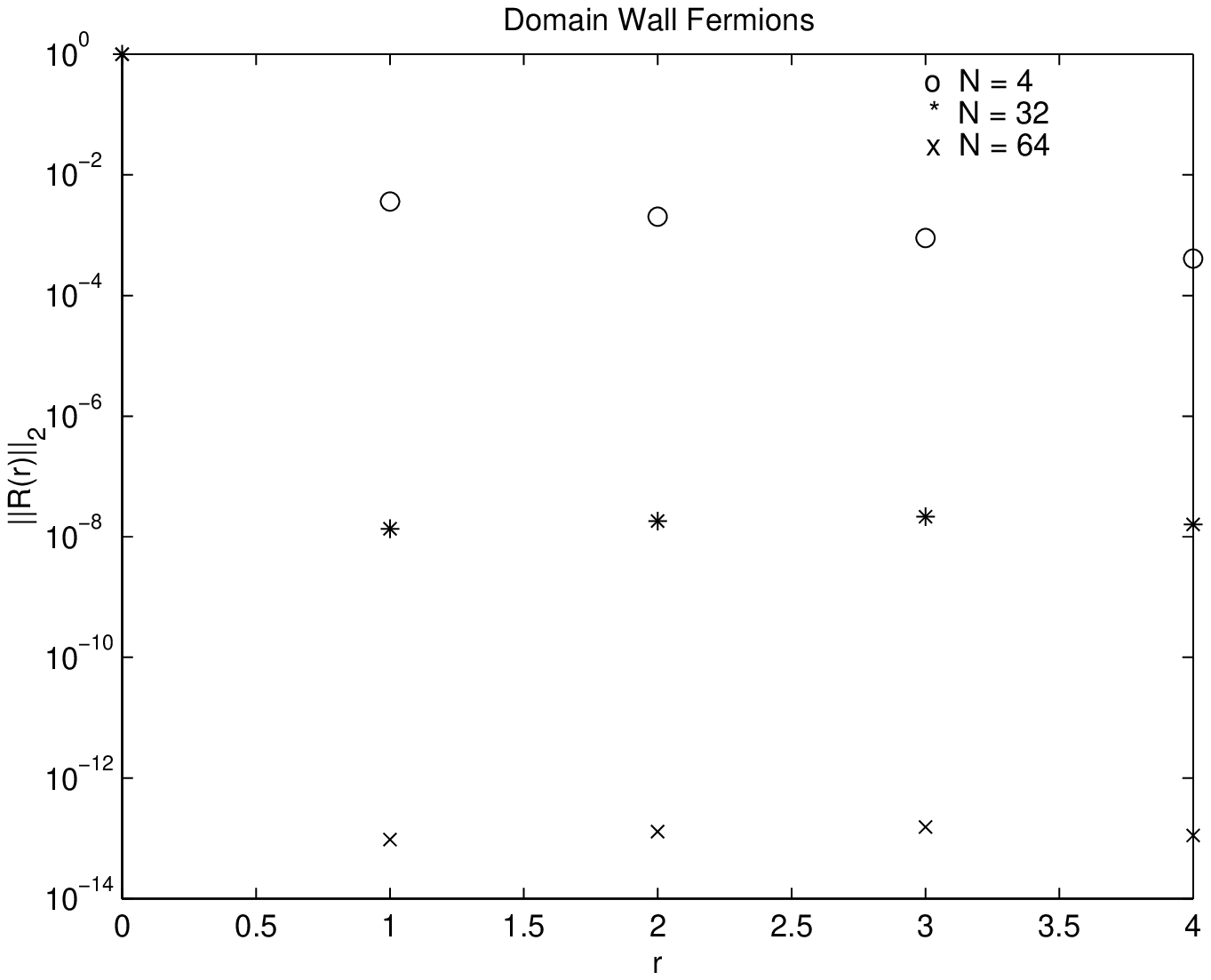}}
\caption{
Norm of the $R$ kernel in spin and color space with the distance $r$ from
the origin for $N = 4$ (circles), $N = 32$ (stars) and $N = 64$ (crosses).
}
\end{figure}
\begin{figure}
\epsfxsize=14cm
\vspace{3cm}
\centerline{\epsffile[100 200 500 450]{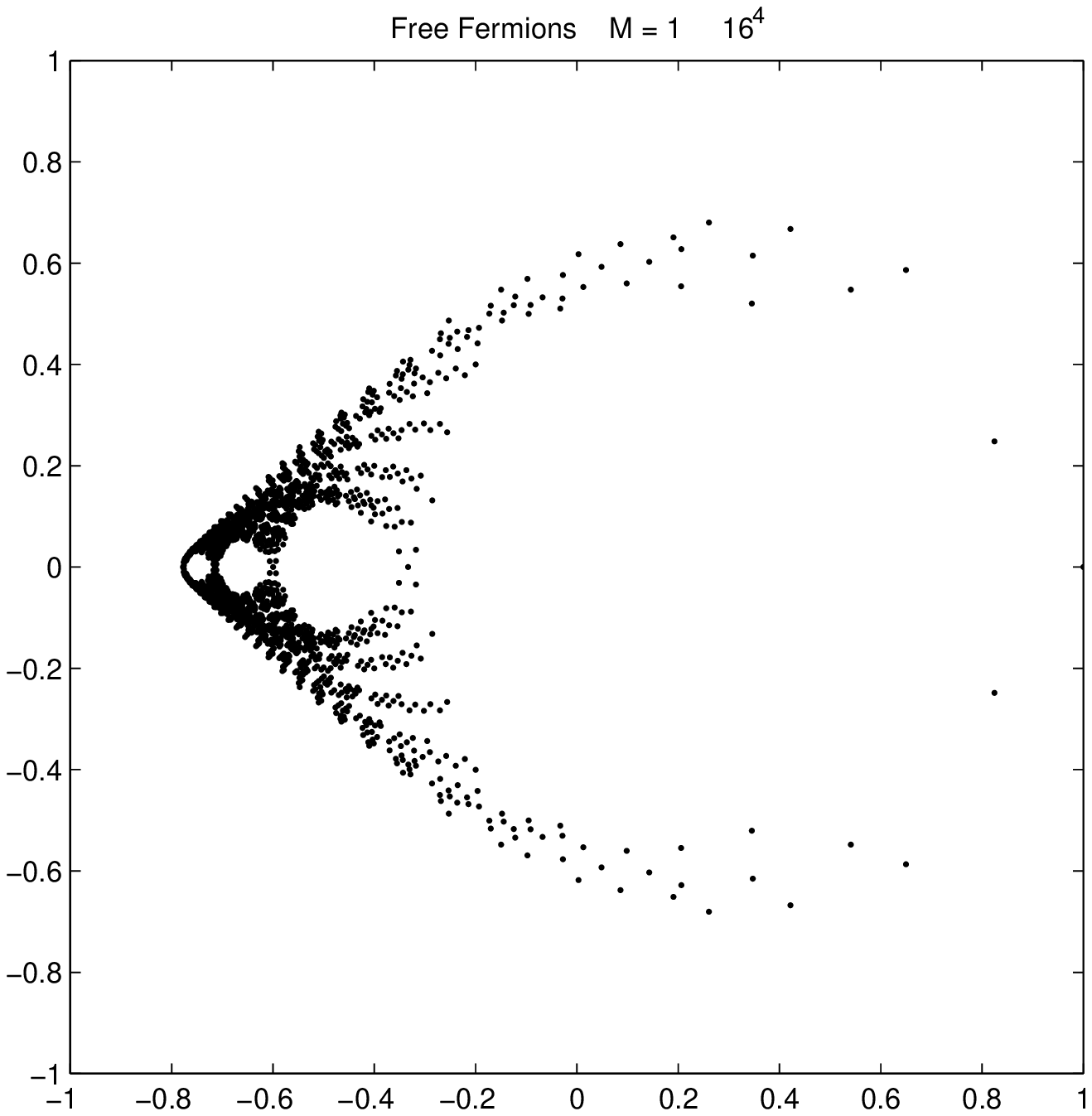}}
\caption{
The spectrum of the $D^{||}/(2 - D^{||})$ matrix in the complex plane.
}
\end{figure}
\begin{figure}
\epsfxsize=8cm
\vspace{3cm}
\centerline{\epsffile[100 200 500 450]{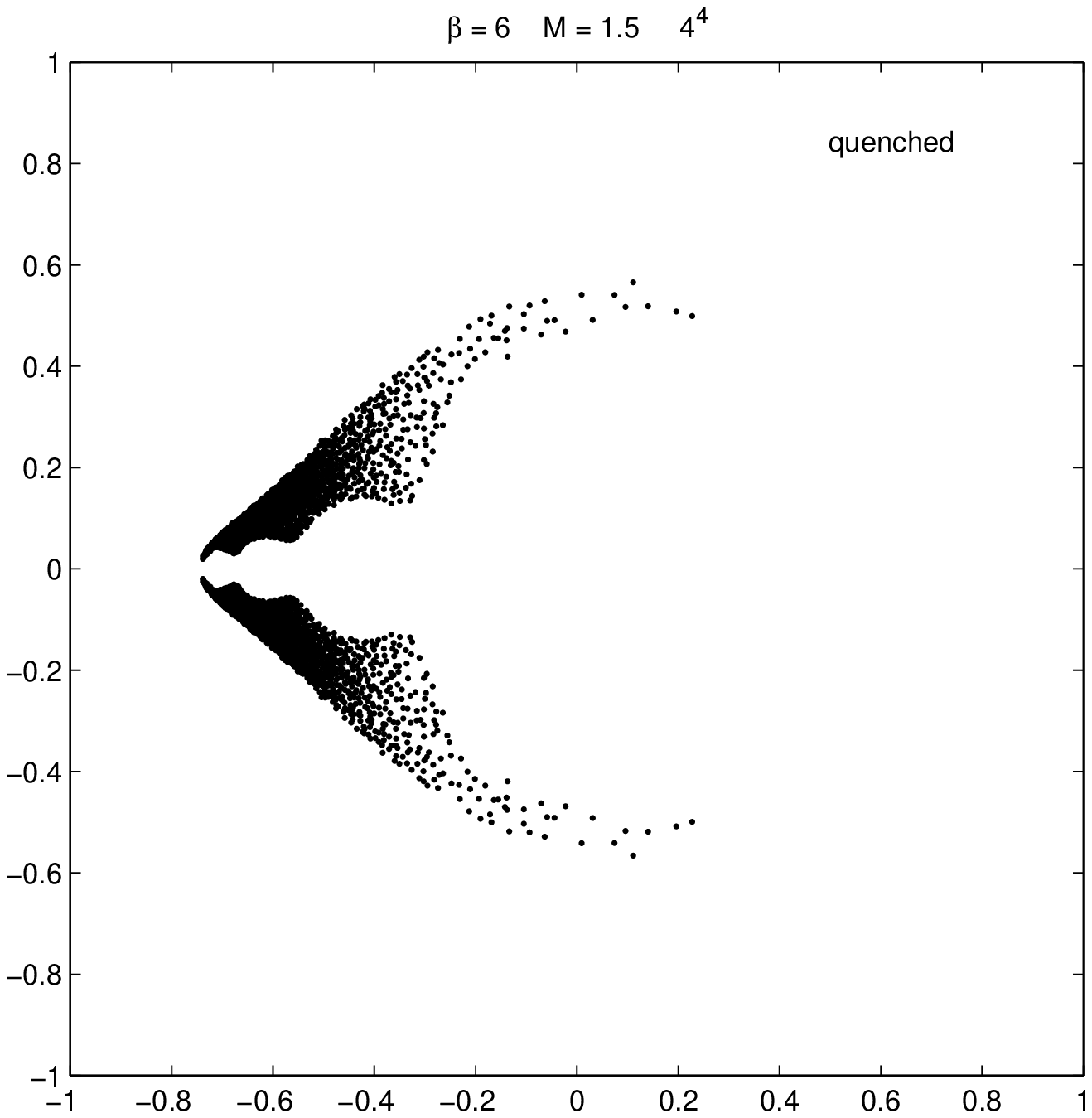}}
\caption{
The spectrum of the $D^{||}/(2 - D^{||})$ matrix in the complex plane.
}
\end{figure}
\begin{figure}
\epsfxsize=8cm
\vspace{3cm}
\centerline{\epsffile[100 200 500 450]{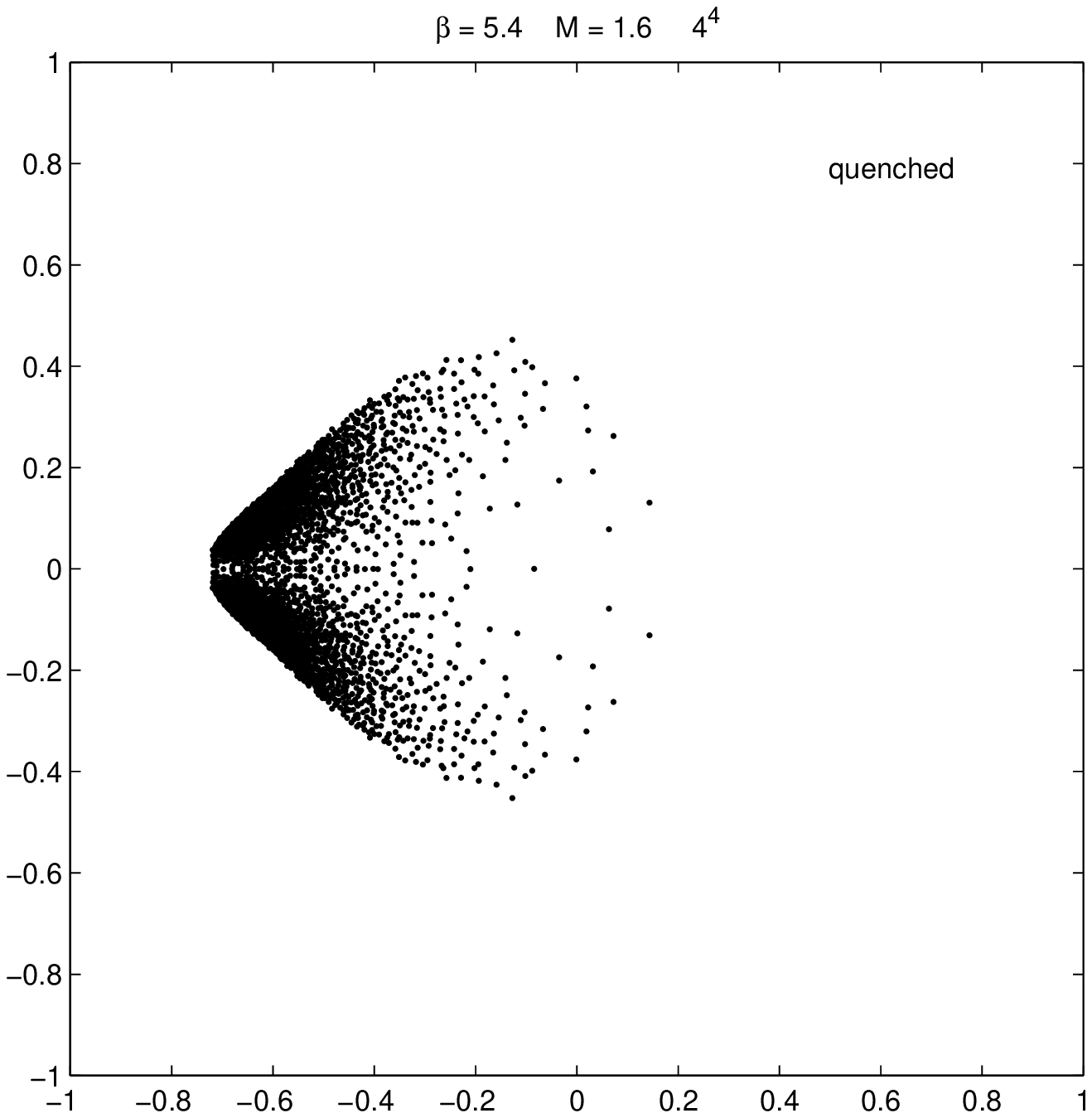}}
\caption{
The spectrum of the $D^{||}/(2 - D^{||})$ matrix in the complex plane.
}
\end{figure}
\begin{figure}
\epsfxsize=12cm
\vspace{3cm}
\centerline{\epsffile[100 200 500 450]{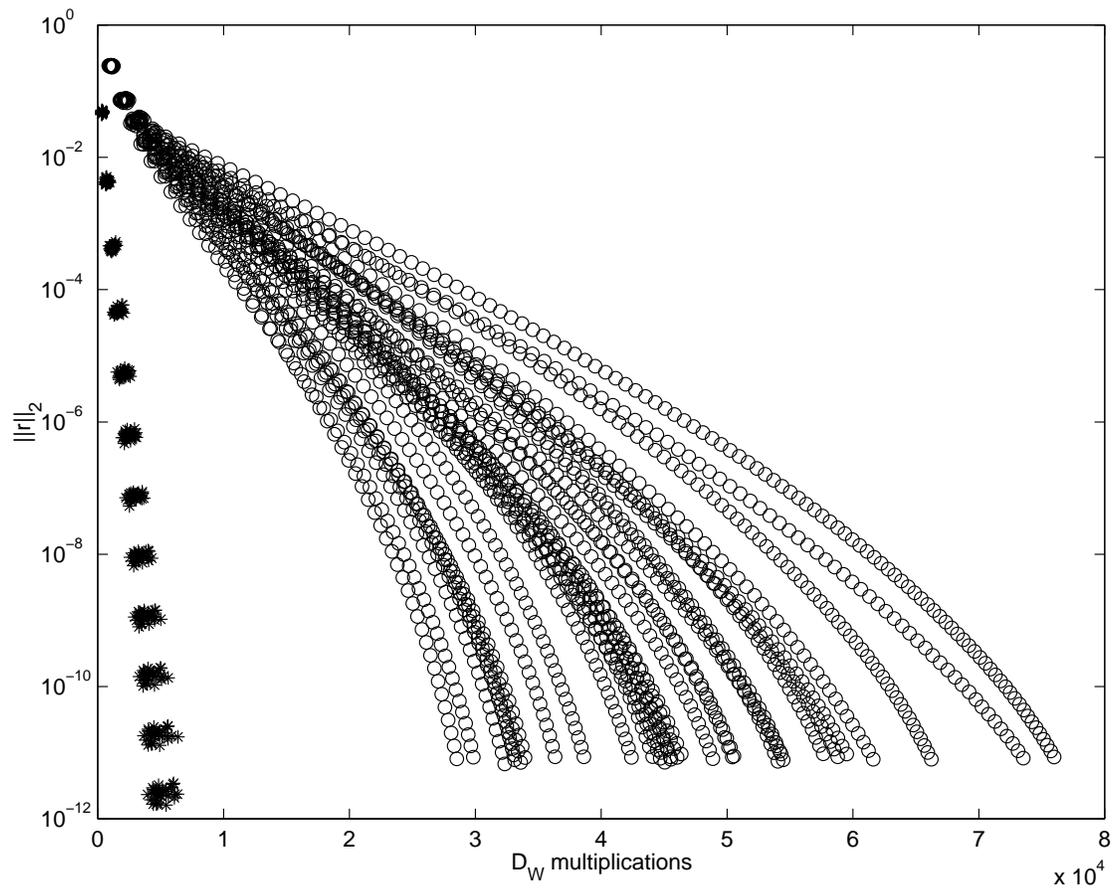}}
\caption{Norm of the residual error vs. the number of $D_W$
multiplications on 30 configurations. Circles stand
for the straightforward inversion with CR and stars for the
multigrid algorithm.}
\end{figure}

\end{document}